# In Situ Cryomodule Demagnetization

Anthony C. Crawford     Fermilab Technical Div. / SRF Development Dept.     acc52@fnal.gov     23Jul15

The feasibility of in-situ demagnetization of fully assembled superconducting RF cryomodules is demonstrated. Useful parametric values for demagnetization as well as measured effects on sensitive components within the cryomodule are listed and discussed. A practical arrangement for active compensation of the axial component of magnetic field is described.

## Introduction

This note is the last in a series that addresses the problems involved with magnetic shielding of cryomodules that are made up of multicell elliptical superconducting cavities. A basic understanding of the various problems is assumed on the part of the reader. For a full contextual understanding of the material contained herein, the reader is advised to see references [1] through [3]. The goal of this and preceding studies is to achieve a practically realizable cryomodule where the cavities are exposed to an average magnetic field of 5 milliGauss or less during cooldown through the normal to superconducting transition temperature of 9.2K. The focus is on details of internal components of the cryomodule and how they affect and are affected by the demagnetization process.

## The Cryomodule

A Fermilab style International Linear Collider (ILC) cryomodule, as shown in Figure 1, has been used for this demagnetization study. The internal parts consist of one helium gas return pipe, eight cavity helium vessels, eight single layer Cryoperm magnetic shields, one quadrupole magnet steel assembly and one cavity tuner step motor with its associated harmonic drive assembly. The steel vacuum vessel had been previously demagnetized as described in reference [3]. Individual cavity Cryoperm shields are open ended for this study with the exception of the shield that was adjacent to the location of the quadrupole magnet. This shield was fitted with cylindrical endcap assemblies. Results of measurements and calculations on the ILC cryomodule will be used to predict behavior for Linac Coherent Light Source-II (LCLS2) cryomodules.

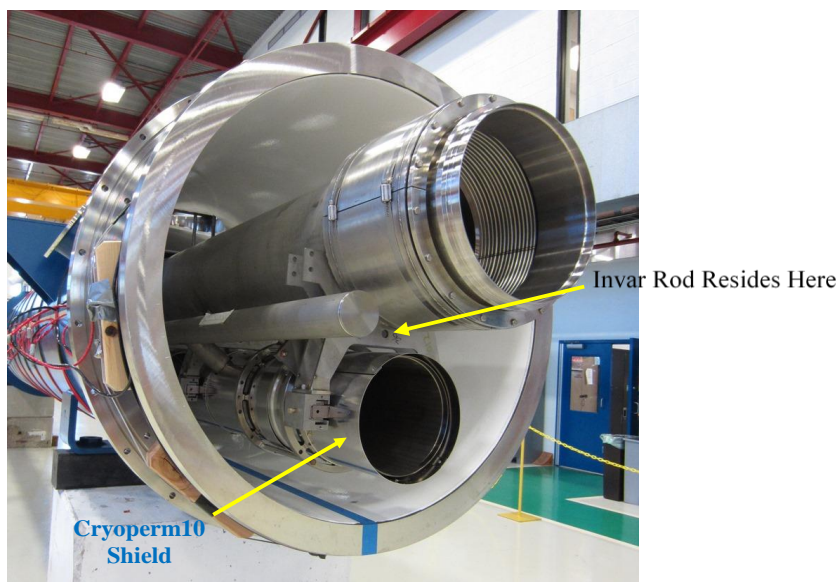

Figure 1.     The Partially Assembled Cryomodule

The cryomodule was demagnetized by exciting a series of coaxial coils with a decreasing bipolar waveform. The coils are shown in Figure 2.

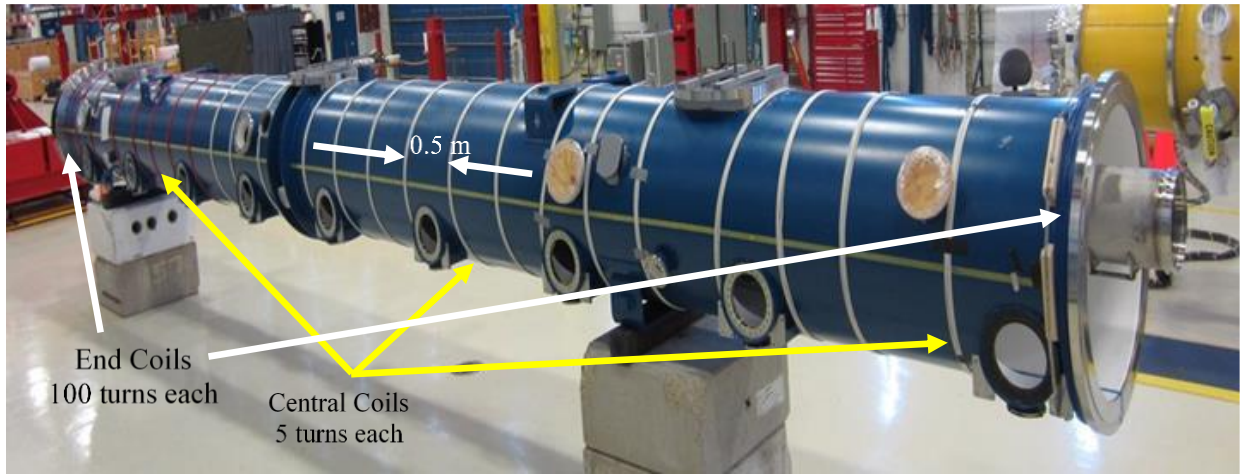

Figure 2.   Coils used for Demagnetization

### The Demagnetization Procedure

The peak magnetomotive force density used for demagnetizing the central 90% of the cryomodule was 650 Ampere-turns/meter. Additional turns were used at the ends of the cryomodule in order to keep the magnetic field in the steel pipe from dropping by more than 10% relative to the value in the center of the cryomodule. The coil winding density for the central portion was 10 turns per meter. Each end coil consisted of 100 turns. All coils were connected in series with their fields aligned in the same direction for the demagnetization procedure. It is intended that the coils will be a permanent part of the cryomodule and will serve the dual purpose of demagnetization and active cancellation of the axial component of the magnetic field when the cryomodule is located in the LCLS2 linac.

The demagnetization current waveform is shown in normalized form in Figure 3. The amplitude of the flat top decrements by 1% of the initial current value per cycle and continues for 100 cycles. The goal is to achieve a demagnetization hysteresis curve similar to the idealized curve shown in Figure 4.

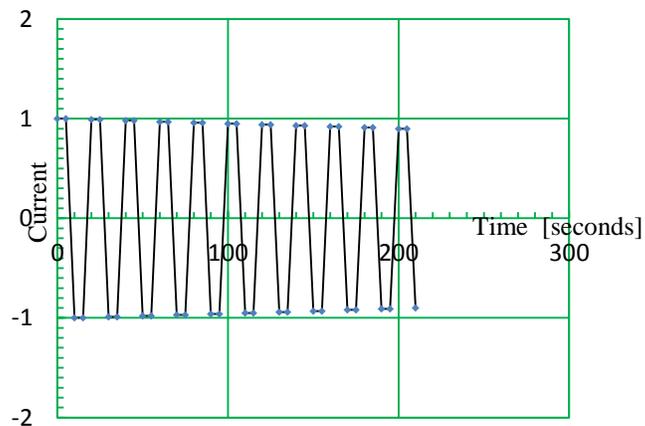

Figure 3.   The Waveform Used for Demagnetization

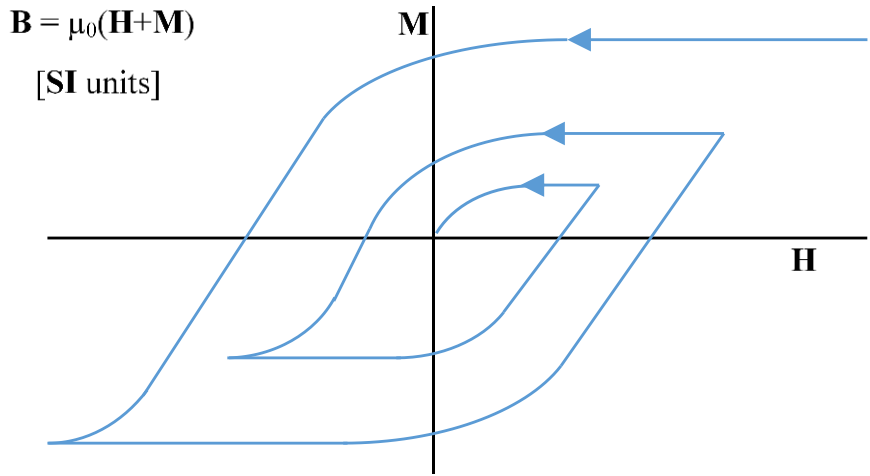

Figure 4.   A Hysteresis Curve for Demagnetization

## Calculations

The anticipated relative initial permeability of Cryoperm at different temperatures is shown in Table 1. This information forms the basis of argument for the relevance of room temperature measurements in predicting magnetic field attenuation when the cavities are at 9.2K. The use of the value $\mu_r = 12{,}000$ allows direct comparison to calculated values in the Tesla Test Facility design report of 1995 [1], p162. A constant relative permeability of 500 was assumed for the steel pipe. Details of the **B**-**H** curve for Cryoperm, after installation in a cryomodule, are not known at this time.

| | |
|---|---|
| Cryoperm $\mu_r$ at 300K (measured) | 10,200 |
| Cryoperm $\mu_r$ at 4K (measured) | 13,300 |
| Cryoperm $\mu_r$ Assumed for Calculations | 12,000 |

Table 1.   Cryoperm Properties

A two dimensional finite element model was used to estimate the magnetic field level inside the steel vacuum vessel and the Cryoperm shields. At an excitation of 650 Ampere-turns/m, the field in the steel is 0.35 Tesla ± 0.07 Tesla. The field is plotted in Figure 5.

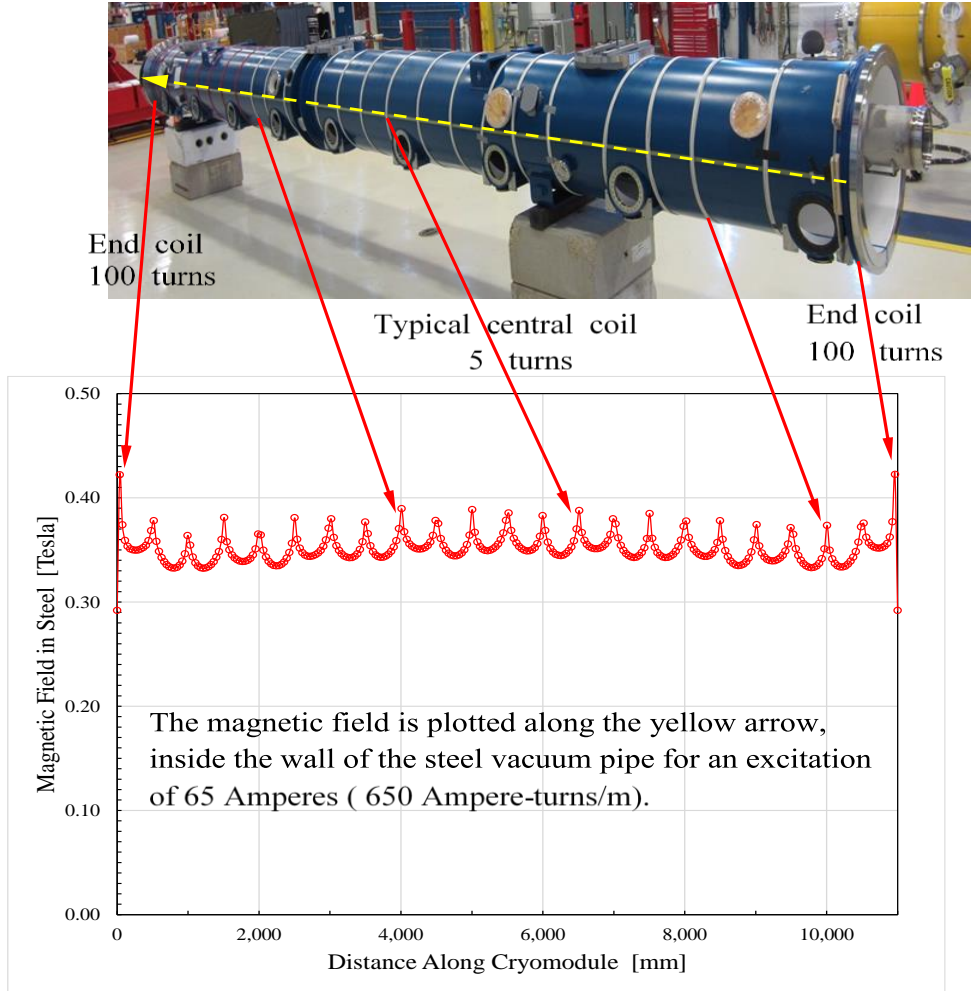

Figure 5.　The Magnetic Field Inside the Wall of the Steel Pipe

With the same 650 Ampere-turns/m excitation, the Cryoperm shields have B >0.4 Tesla near the midpoint of their length.  The ends of the Cryoperm shields have lower fields than their central portions.
　　The magnetic field to which the Invar rod, tuner motors and the quadrupole magnet will be exposed during a 650 Ampere-turn/m demagnetizing procedure was calculated.  Results are listed in Table 2.

| Component | Calculated Ambient Field at 650 Ampere-turns/m Excitation |
|---|---|
| Invar Rod | 0.0002 Tesla |
| Tuner Step Motor | 0.003 Tesla |
| Quadrupole Magnet | 0.005 Tesla |

Table 2.　The Calculated Ambient Field Experienced by Sensitive Components

### Measurement Technique

　　The effect of de-magnetization was measured by sampling the axial magnetic field component along the centerline of the Cryoperm shield. This line coincides with the cavity and beamline axis in a cryomodule. All measurements were taken at room temperature. Resolution and accuracy of the field measurements was better than ± 0.5 milliGauss. Although measurement along the cavity axis is a good way to judge the effect of remnant fields on SRF cavity performance, it does not lead to a detailed understanding of the demagnetization process for individual components. The remnant field of individual cryomodule components was not measured for this study.

## Results

### The Minimum Excitation Required to Change the Remnant Field

Results of demagnetization with initial excitation current greater than 20 Amperes ( 200 Ampere-turns/m) indicate that the alternating current waveform should not be terminated until the current is significantly lower than 200 milliAmperes. No remnant magnetic effects along the axis of the Cryoperm shields could be discerned following coil excitation values below 200 milliAmperes (~ 2 Ampere-turns/meter). After excitation at levels above this value, the field inside the Cryoperm shield would change. Because of this, the length of the demagnetization waveform was extended so that the final cycle was at 0.1 Ampere-turns/meter. To achieve demagnetization with a bipolar waveform, it is necessary that there be no DC offset current that approaches the magnitude of 200 milliAmperes. The power supply used for this study was adjusted to have DC offset less than 10 milliAmperes. Temperature stabilization of the power supply regulation circuitry may be required to keep the offset within tolerances for cases where the power supply is located in an unfavorable environment.

If a constant relative permeability value of 500 is assumed for the steel used for the vacuum vessel, and a constant relative permeability of 12,000 is assumed for the Cryoperm, then a 2 Ampere-turns/m excitation results in approximately 11 Gauss inside the wall of the steel pipe and in a maximum of approximately 27 Gauss inside the wall of the Cryoperm shields. Implications of this observation are developed in the "Discussion" section of this note.

### The Effect of Demagnetization on the Tuner Step Motor

There has been concern over the effect of cryomodule demagnetization on the permanent magnet based motors that operate the cavity frequency tuners. The technique used to judge whether the motor has been degraded by demagnetization of the cryomodule was to measure the torque generated by the motor at a prescribed current both before and after the demagnetization. No change in the torque value was observed up to the full excitation of 650 Ampere-turns/m. It is therefore considered safe to demagnetize the cryomodule with the step motors in place.

### The Effect of Demagnetization with the Invar Rod and Quadrupole Magnet in Place

The presence of these magnetic components had no harmful effect on the demagnetization. The axial field component along the centerline of the Cryoperm shield closest to the quadrupole is shown in Figure 6, both before and after demagnetization. During the demagnetization and the subsequent measurements, the quadrupole magnet was located at the distance from the first Cryoperm shield that it will have in LCLS2 cryomodules. The end of the quadrupole steel is located at the x-axis value of -41 cm in Figure 6. The magnet steel used for this test was the LCLS2 quadrupole design.

The cryomodule was intentionally magnetized for the "Before" data shown in Figure 6. This was done with a 10 Ampere-turn/m excitation of the coils. The magnetization procedure was necessary because the cryomodule had been previously demagnetized for earlier studies. The field along the cavity axis is dominated by remnant field in cryomodule components in the "Before Demagnetization" data of Figure 6.

The following facts are relevant to the field measurement shown in Figure 6. No axial field cancellation was used. The ambient environmental axial field for the cryomodule was approximately 50 milliGauss, the location and orientation of the cryomodule having been chosen to minimize this value. The coordinates 0 through 100 centimeters coincide with the location of the active length of the cavity that would be present in a complete cryomodule cold mass assembly. An extremely short (2 cm) endcap was used at the end of the shield near 100 cm, explaining the strong increase in field at this end of the cavity.

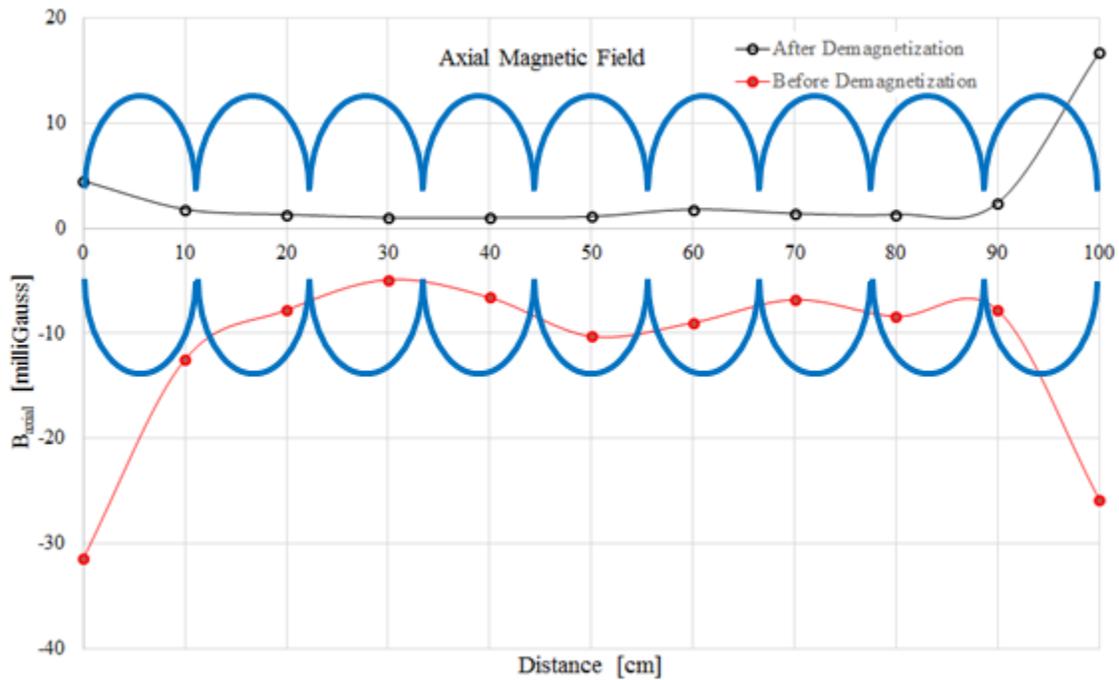

Figure 6.   The Axial Component of Field Inside the Cryoperm Shield Nearest to the Quadrupole Magnet Before and After Demagnetization  ( The profile of the cavity is superimposed on the longitudinal coordinate axis to show the relative position of the cells.)

The question of whether the superconducting quadrupole magnet for the LCLS2 cryomodule will create remnant field in the vacuum vessel steel and the Cryoperm shields has been addressed elsewhere [4].  Remnant field in nearby cryomodule components caused by the quadrupole at full excitation current will have an insignificantly small effect at the location of the nearest cavity cell.

The effect of stray remnant field in the steel flux return of the LCLS2 quadrupole magnet has also been measured.  One quadrupole magnet assembly that had previously been cold tested to its maximum current was placed in the cryomodule at a distance of 41 centimeters from the first cavity cell.  The steel of the magnet flux return was magnetized to a larger value than is anticipated in cryomodule use, i.e. up to the quench current of the magnet.  The magnetic field was measured along the cavity centerline inside the Cryoperm shield with the quadrupole in place.  The magnet assembly was then removed from the cryomodule and the magnetic field along the cavity centerline re-measured.  The result was that no change in the field at the cavity was observed that was larger than ± 0.5 milliGauss.  Stray field from the quadrupole magnet assembly is judged to be acceptable and it is unlikely that demagnetization of the quadrupole flux return will be necessary.

## Discussion

The observation that a 2 Ampere-turns/m excitation can re-arrange magnetic domains within the cryomodule has implications about how demagnetization and cancellation of axial fields must be handled.  In particular, the central section of the steel pipe will experience a peak internal field of ~200 Gauss should it be exposed to a uniform ambient axial field equal to 500 milliGauss.  This will, in principle, lead to a significantly large remnant field inside the steel.   Uniform environmental fields of this magnitude are possible during transport and handling. This means that, for the LCLS2 project, cryomodule demagnetization is likely to be required both before initial tests at Fermilab and JLab and after the cryomodules are installed in their final operational locations at SLAC.

After demagnetization, magnetic domains within the cryomodule will re-arrange themselves according to the local magnetic environment. The demagnetization coils can then be used to minimize the average axial field inside the Cryoperm shields when used in conjunction with fluxgate magnetometers aligned with the axial direction. Flux gate sensors are chosen because they are capable of measuring absolute field magnitudes on the order of 1 milliGauss without the necessity of periodic zero point offset and scale factor adjustments. Specialized models of fluxgate magnetometers [5] are reliable in cryogenic applications with very low error introduction over the temperature range 2K to 300K.

The preferred choice for fluxgate location is on the cavity beamtubes within the innermost layer of magnetic shielding. This is because not only can the sensors be used to adjust the axial field cancellation coils, but in addition, valuable information on cavity flux expulsion during normal state to superconducting state transition can be acquired. Also, the axial magnetic field component of Seebeck coefficient induced thermal current can be observed during the cooldown process [6]. Transient axial magnetic fields on the order of tens of milliGauss have been observed during cavity cooldown procedures in horizontal tests at Fermilab.

The proposed arrangement of fluxgate sensors for LCLS2 axial field cancellation is shown in Figure 7. Fluxgate positions are labelled FG1 through FG5. Cavity positions are labelled C1 through C8. Fluxgates FG2, FG3 and FG4 will be used to adjust the central cancellation coil current. Fluxgates FG1 and FG5 will be used to adjust the two independent trim coils.

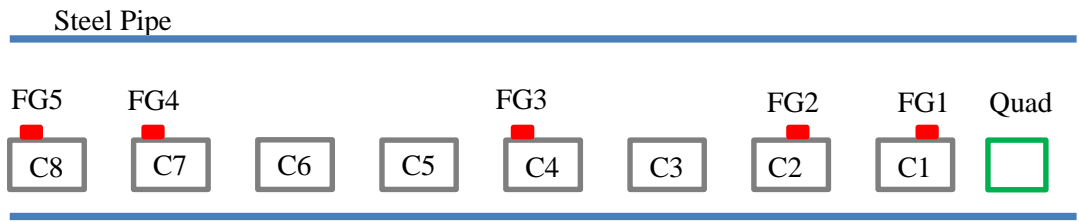

Figure 7. Fluxgate Magnetic Sensor Positions

LCLS2 cavity magnetic shields consist of two concentric cylindrical layers of one millimeter thick Cryoperm with a radial separation of two centimeters between the layers. In the most recent version of the shield design, the inner layer of shields and end caps are designed to fit as closely as possible to the cavity helium vessel and beamtubes. This means that there is no available space for fluxgates inside the shield layer. However, placing the fluxgates in the two centimeter radial gap between the two layers of Cryoperm serves as a suitable compromise for the purpose of cancellation coil adjustment. A graph of calculated axial field at the fluxgate locations between the shield layers is shown along with the field along the cavity axis in Figures 8 and 9. The graphs in Figures 8 and 9 were calculated using a two dimensional finite element model.

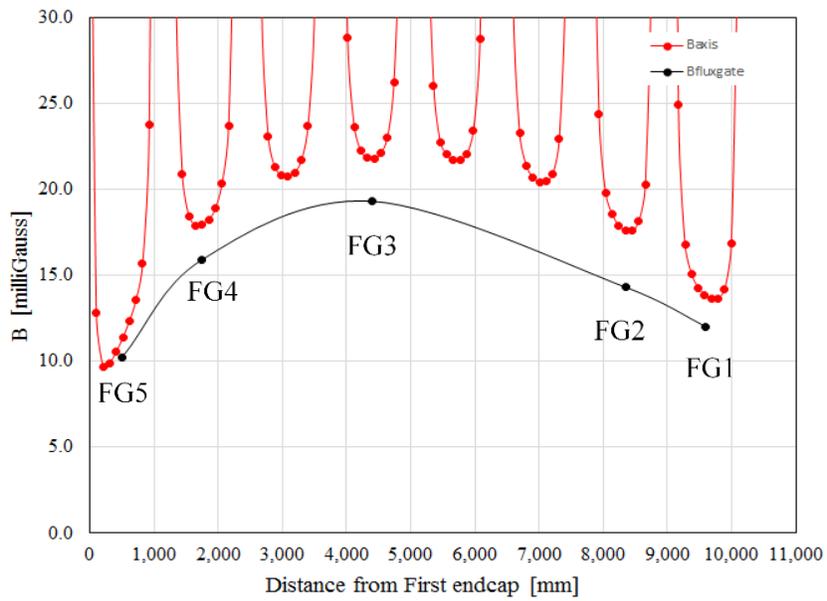

Figure 8.  Fluxgate Magnetic Sensor Readings with Cancellation Coils Off, $B_{axial}$ Ambient = 200 milliGauss

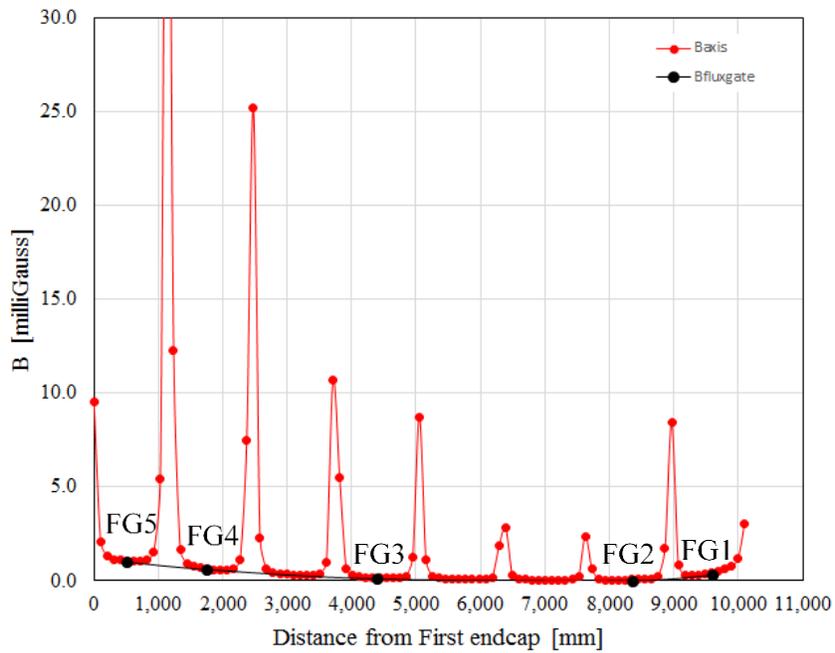

Figure 9.  Fluxgate Magnetic Sensor Readings with Cancellation Coils On, $B_{axial}$ Ambient = 200 milliGauss

The axial field as sampled in the gap between the shielding layers is adequate for minimizing the longitudinal field component, although information about flux dynamics during cavity cooldown will be forfeited for this arrangement.

**Related Subjects**

Thermal current during cryomodule cooldown is both interesting and complex. It is probably not a good practice to attempt to minimize thermal current readings by means of adjusting cancellation coils. The purpose of the cancellation coils is to minimize the magnitude of field that the Cryoperm shields are to attenuate. Thermal currents that result in an increase in the axial field component are likely to be local effects that cannot be effectively cancelled with magnetic correctors of large spatial extent without increasing the field in other locations.

It is also interesting to consider the use of counter magnetization of the cryomodules to cancel the local axial field component. The technique involves applying a sufficiently large DC excitation to the demagnetization coils to overcompensate for the local axial field to the extent that when the current is set to zero, the field measured by the fluxgates is minimized. This has been done successfully for relatively low ambient axial field ( ~ 100 milliGauss ) with the Fermilab Cryomodule assembly. Counter magnetization potentially removes the necessity for active compensation and lowers risk due to malfunctioning or incorrectly adjusted DC power supplies.

One demagnetizing scenario that is planned for the initial LCLS2 cryomodule test at Fermilab is to allow the cryomodule to reach thermal equilibrium with liquid helium temperature equal to 2K, then perform a demagnetization procedure followed by adjustment of the cancellation coils. The cavities would then be warmed to a temperature above 20K and then cooled rapidly below 9.2K. This procedure would, in principle, provide a demagnetized cryomodule with optimally cancelled axial field, all adjusted reasonably close to the niobium transition temperature of 9.2K.

Is it necessary to demagnetize the steel vacuum pipe prior to the assembly of the cryomodule? The answer is "Probably not." A preliminary remnant field measurement would identify any part of the empty steel pipe that had exceptionally large remnant field and that needed special attention. To date, the maximum remnant field levels measured at the surface of the pipe have been at the level of 3 Gauss and can be reset by the demagnetization procedure described here for a fully assemble cryomodule. More information on this subject will be collected during the first stages of cryomodule production for LCLS2.

**References**

[1] The Conceptual Design Report for the TESLA Test Facility Linac, Version 1.0
http://tesla.desy.de/TTF_Report/CDR/pdf/cdr_chap4.pdf, p.162

[2] Crawford, A., "A Study of Magnetic Shielding Performance of a Fermilab International Linear Collider Superconducting RF Cavity Cryomodule" http://arxiv.org/abs/1409.0828

[3] Crawford, A. "Superconducting RF Cryomodule Demagnetization",
http://arxiv.org/ftp/arxiv/papers/1503/1503.04736.pdf

[4] Terechkine, I., "Fringe Field of a Focusing Quadrupole in the Beamline of the LCLS Cryomodule",
https://web.fnal.gov/organization/TDNotes/Shared%20Documents/2014%20Tech%20Notes/TD-14-005.pdf

[5] http://www.bartington.com/literaturePDF/application%20notes/AN0029%20Mag-01H%20used%20for%20Field%20Measurements%20in%20Cryogenic%20Chambers.pdf

[6] Crawford, A. "A Study of Thermocurrent Induced Magnetic Fields in ILC Cavities"
http://arxiv.org/ftp/arxiv/papers/1403/1403.7996.pdf